\documentclass[prb, twocolumn,amsmath,amssymb,superscriptaddress,floatfix]{revtex4}

\usepackage{graphicx}
\usepackage{dcolumn}
\usepackage{bm}
\usepackage{color}

\begin{document}


\title{
Spatial gradient of dynamic nuclear spin polarization \\
induced by breakdown of quantum Hall effect
}

\author{M. Kawamura}
	\email{minoru@riken.jp}
	\affiliation{RIKEN Advanced Science Institute, Wako, Saitama 351-0198, Japan}
	\affiliation{PRESTO, Japan Science and Technology Agency, 
		Kawaguchi, Saitama 333-0012, Japan}
\author{K. Kono}
	\affiliation{RIKEN Advanced Science Institute, Wako, Saitama 351-0198, Japan}
\author{Y. Hashimoto}
	\affiliation{Institute for Solid State Physics, University of Tokyo, 
		Kashiwa, Chiba 277-8581, Japan}
\author{S. Katsumoto}
	\affiliation{Institute for Solid State Physics, University of Tokyo,
		Kashiwa, Chiba 277-8581, Japan}
	\affiliation{Institute for Nano Quantum Information Electronics, University of Tokyo, \\
					Meguro-ku, Tokyo 153-8505, Japan}
\author{T. Machida}
	\email{tmachida@iis.u-tokyo.ac.jp}
	\affiliation{Institute for Nano Quantum Information Electronics, University of Tokyo, \\
					Meguro-ku, Tokyo 153-8505, Japan}
	\affiliation{Institute of Industrial Science,	University of Tokyo, Meguro-ku, Tokyo 153-8505, Japan}
\date{\today}

\begin{abstract}
We studied spatial distribution of dynamic nuclear polarization (DNP)
in a Hall-bar device in a breakdown regime of the quantum Hall effect (QHE).
We detected nuclear magnetic resonance (NMR) signals from the polarized nuclear spins 
by measuring the Hall voltage $V_{xy}$ 
using three pairs of voltage probes attached to the conducting channel of the Hall bar.
We find that the amplitude of the NMR signal depends on the position of the Hall voltage probes
and that the largest NMR signal is obtained from 
the pair of probes farthest from the electron-injecting electrode.
Combined with results on pump-and-probe measurements,
we conclude that the DNP induced by QHE breakdown develops
 along the electron-drift direction.
\end{abstract}

\maketitle

The current-driven breakdown of the quantum Hall effect (QHE) is a phenomenon,
characterized by an abrupt increase of longitudinal resistance
and a deviation of Hall resistance from the quantized resistance,
when the electric current passing through a  
quantum Hall (QH) conductor exceeds a critical current $I_{\rm c}$\cite{Ebert1983,Cage1983}.
The QHE breakdown has gained renewed interest because of the recent development 
of the dynamic nuclear spin polarization (DNP) technique
in QHE breakdown regimes\cite{Song2000,Kawamura2007,Kawamura2008,Dean2009}.
It was demonstrated that 
nuclear spins in GaAs/AlGaAs heterostructures are dynamically polarized
through the hyperfine interaction between electron spins and nuclear spins
when the two-dimensional electron system (2DES) embedded in the heterostructure 
is driven to the breakdown regime of odd-integer QHE.
Given its simplicity, the DNP technique can be useful for the initialization procedure
of nuclear spin quantum bits\cite{Machida2003,Yusa2005,Takahashi2007}.
Moreover, the back-action of the polarized nuclear spins
on electrical transport coefficients of the 2DES
makes it possible to perform extremely sensitive detection of nuclear magnetic resonance (NMR).
Thus, the resistively detected NMR technique can be utilized to investigate
the electron spin properties of QH systems
\cite{Hashimoto2002,Smet2002,Stern2004,Kumada2007}.

\begin{figure}[btp]
	\begin{center}
	\includegraphics[width=7.8cm]{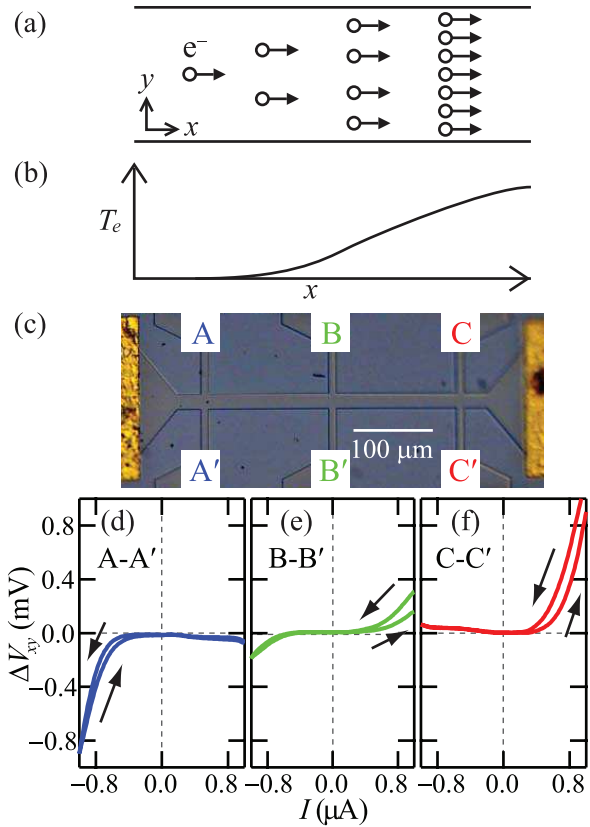}
	\caption{
		\label{IVcurve}
		(a) 
		Schematic of avalanche multiplication of excited electrons
		in a breakdown regime of the quantum Hall effect\cite{Komiyama1996}.
		The circles indicate electrons in the upper Landau subband
		and the arrows indicate the electron-drift direction.
		(b)
		Schematic of spatial gradient of $T_{e}$
		expected in the QHE breakdown regime when electrons
		are injected from the left end.
		(c) Micrograph of the Hall-bar device used in the present study.
		(d)-(f)  $\Delta V_{xy}$-$I$ curves obtained by sweeping current
		 in positive and negative directions
		at $B$ = 8.65 T ($\nu$ = 1.07).
	  	Hall voltage probe pairs (d) A-A$^\prime$, (e) B-B$^\prime$, and (f) C-C$^\prime$
		were used to obtain the data.
		The deviation $\Delta V_{xy}$ is defined by $\Delta V_{xy} = h/e^2 I - V_{xy}$.
		The arrows indicate the directions of the current sweep.
	}
	\end{center}
\end{figure}

Although 
QHE breakdown has been studied\cite{Ebert1983, Cage1983} both theoretically and experimentally
ever since the discovery of QHE,
earlier experimental studies were mainly conducted
at even-integer QHE where the spin degree of freedom is unresolved.
In the case of even-integer QHE,
QHE breakdown does not occur homogeneously in the QH conductor
but develops over a macroscopic distance (typically 100 $\mu$m)
along the electron-drift direction\cite{Komiyama1996,Kaya1999}.
The development of  QHE breakdown over a macroscopic distance
was explained\cite{Komiyama1996, Komiyama2000, Kaya1999,Guven2002}
on the basis of avalanche multiplication of excited electrons in the upper Landau subband:
the number of electrons in the upper Landau subband
increases along the electron-drift direction in the QH conductor.
This means that electron temperature $T_{e}$  increases along
the conducting channel of a Hall-bar device [Figs.~\ref{IVcurve}(a) and (b)].
Such a spatial distribution of $T_{e}$
in  QHE breakdown regimes was directly observed in
resistance measurements in multi-terminal Hall bars\cite{Komiyama1996, Kaya1999, Kaya1998}
and cyclotron emission imaging experiments\cite{Kawano2005}.

In the case of the breakdown of odd-integer QHE,
the electron excitation process is accompanied by electron spin flips.
In our earlier work, we proposed that electron spin flips 
associated with the avalanche multiplication of excited electrons produce  DNP
through hyperfine interaction\cite{Kawamura2007}.
If our scenario is correct, 
the DNP induced by  QHE breakdown should be spatially distributed
over a macroscopic distance
because the avalanche multiplication process produces
a spatial distribution not only of $T_{e}$ but also of electron spin flip events.
However, in  earlier studies\cite{Kawamura2007} on DNP,
little attention was paid to its spatial distribution,
and DNP signals were collected by  longitudinal voltage measurements
in which the signals were averaged over a macroscopic area
between the voltage probes of the Hall bar.
So far, no experimental result has been reported
concerning the spatial distribution of  DNP.
In order to reveal the spatial distribution of  DNP, 
it is necessary to detect DNP signals locally in the Hall bar.

In this paper, 
we report that the DNP
induced by QHE breakdown is spatially   distributed in a Hall-bar device.
We detected NMR signals from the polarized nuclear spins 
by measuring the Hall voltage $V_{xy}$ 
using three pairs of voltage probes attached to the conducting channel
of the Hall bar.
Since  the Hall voltage $V_{xy}$ reflects local electronic properties
at the intersection of a conducting channel and a pair of voltage probes\cite{Geim},
the spatial distribution of DNP can be resolved by measurements of $V_{xy}$.
We find that the amplitude of the NMR signal depends on the position of the Hall voltage probes
and that the largest NMR signal is obtained from 
the pair of Hall voltage probes farthest from the electron-injecting electrode.
Combined with results on pump-and-probe measurements, 
we conclude that the DNP induced by QHE breakdown develops
along the electron-drift direction.

We used a GaAs/Al$_{0.3}$Ga$_{0.7}$As single heterostructure wafer 
with a 2DES at the interface.
The mobility and sheet carrier density of the 2DES at 4.2 K were
$\mu$ = 107 m$^2$V/s and $n$ = 2.23 $\times$ 10$^{15}$ m$^{-2}$, respectively. 
The wafer was processed into a 20-$\mu$m-wide Hall bar as shown in Fig.~\ref{IVcurve}(c).
Three pairs of Hall voltage probes
were attached to the conducting channel of the Hall bar, each separated by 155 $\mu$m.
The Hall-bar device was cooled down to a temperature of $T$ = 1.5 K,
and a magnetic field of $B$ was applied perpendicular to the 2DES.
A single-turn coil wound around the device was used to
irradiate a rf-magnetic field of $B_{\rm rf}$ for NMR measurements.

Figures~\ref{IVcurve}(d)-(f) show the dependence of 
 $\Delta V_{xy}$, which is defined by $\Delta V_{xy} \equiv h/e^2 I - V_{xy}$,
on the  bias current $I$.
When $I$ is smaller than the critical current $I_{\rm c}$,
the Hall resistance $R_{\rm H} = V_{xy}/I$  is quantized to $h/e^2$
and $\Delta V_{xy}$ is zero.
When $I$ exceeds $I_{\rm c}$,
$R_{\rm H}$  deviates from the quantized value to
approach the classical Hall resistance $B/ne$
and $\Delta V_{xy}$ is expected to increase.
Thus, $\Delta V_{xy}$ is a good quantity to probe QHE breakdown in a local region
at the intersection of the conducting channel and the voltage probes.
The data in Figs.~\ref{IVcurve}(d)-(f) were obtained
 by sweeping $I$ up and down at a rate of 13.2 nA/s
using the Hall voltage probes (d) A-A$^\prime$, (e) B-B$^\prime$, and (f) C-C$^\prime$
with $B$ = 8.65 T ($\nu$ = 1.07)\cite{fillingfactor}.
On the positive current polarity side
where electrons are injected from the left electrode in Fig.~\ref{IVcurve}(c),
the value of $\Delta V_{xy}$ at probe C-C$^\prime$
increases sharply as $I$ exceeds a critical current
of about $I_{\rm c}$ = 0.2 $\mu$A [Fig.~\ref{IVcurve}(f)].
The value of $\Delta V_{xy}$ at probe B-B$^\prime$ [Fig.~\ref{IVcurve}(e)] also increases 
above $I_{\rm c}$,
although the change in  $\Delta V_{xy}$ is smaller than that at  probe C-C$^\prime$.
The change in $\Delta V_{xy}$ at probe A-A$^\prime$[Fig.~\ref{IVcurve}(d)] is almost zero.
On the negative current polarity side, 
the above described trends are spatially reversed with respect to the center of the Hall bar:
the change in $\Delta V_{xy}$ is  largest at  probe A-A$^\prime$
and almost zero at  probe C-C$^\prime$ when the current polarity is negative.
These systematic differences in the $\Delta V_{xy}$-$I$ curves 
show that QHE breakdown is more prominent in the region farther from the 
electron-injecting electrode.
The origin of this dependence of the $\Delta V_{xy}$-$I$ curves on the probe position
can be attributed to the spatial gradient of $T_{e}$
in the QHE breakdown regime, similar 
to the earlier studies at even-integer filling factors\cite{Komiyama1996, Kaya1999}.
In addition, we observe hysteresis in the $\Delta V_{xy}$-$I$ curves,
which is a signature of the DNP induced by  QHE breakdown\cite{Kawamura2007}.

\begin{figure}[btp]
	\begin{center}
	\includegraphics[width=8cm]{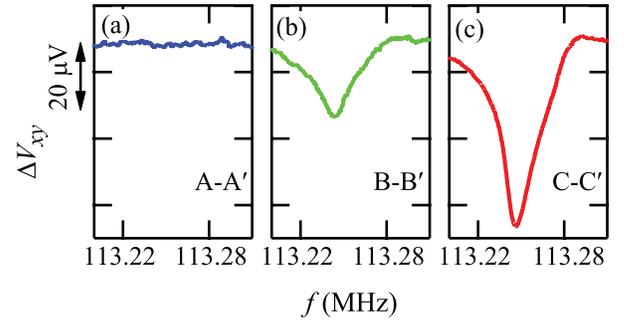}
	\caption{
		\label{spectrum}
		Responses of $\Delta  V_{xy}$ to irradiation of a rf-magnetic field
	 	obtained at $B$ = 8.65 T ($\nu$ = 1.07) and $I$ = $+$0.80 $\mu$A,
		using the Hall voltage probes (a) A-A$^\prime$, (b) B-B$^\prime$, and (c) C-C$^\prime$.
		The frequency of rf-magnetic field was scanned between 113.20 MHz and 113.31 MHz.
		The value of $\Delta V_{xy}$ decreases at an NMR frequency of $^{71}$Ga.
		}
	\end{center}
\end{figure}

The spatial distribution of DNP was investigated
by measuring NMR signals using the  Hall voltage probe pairs
A-A$^\prime$, B-B$^\prime$, and C-C$^\prime$.
Since  the Hall voltage $\Delta V_{xy}$ reflects local electronic properties
at the intersection of the conducting channel and the pair of voltage probes\cite{Geim},
the spatial distribution of DNP can be resolved
by comparing the amplitudes of NMR signals obtained from each pair of Hall voltage probes.
The data shown in Figs.~\ref{spectrum}(a)-(c) were obtained at $B$ = 8.65 T
according to the following procedure.
First, nuclear spins were polarized by applying a current of $I$ = $+$0.8 $\mu$A
through  hyperfine interaction\cite{Kawamura2007}.
Subsequently, $V_{xy}$ was measured under irradiation of a continuous wave 
rf-magnetic field of $B_{\rm rf}$.
As the frequency of $B_{\rm rf}$ was scanned,
the value of $V_{xy}$ decreased
when the frequency matched the NMR frequency of $^{71}$Ga 
(gyro-magnetic ratio $\gamma$ = 81.58 rad$\cdot$MHz/T)
as shown in Figs.~\ref{spectrum}(b) and (c).
The amplitude of the NMR signal is the largest at probe C-C$^\prime$,
intermediate at probe B-B$^\prime$,
and zero  at probe  A-A$^\prime$.
The dependences of the NMR spectrum amplitude 
for $^{69}$Ga  ($\gamma$ = 64.21 rad$\cdot$MHz/T) 
and $^{75}$As ($\gamma$ = 45.82 rad$\cdot$MHz/T) show
a trend similar to that of $^{71}$Ga.

\begin{figure}[btp]
	\begin{center}
	\includegraphics[width=8.5cm]{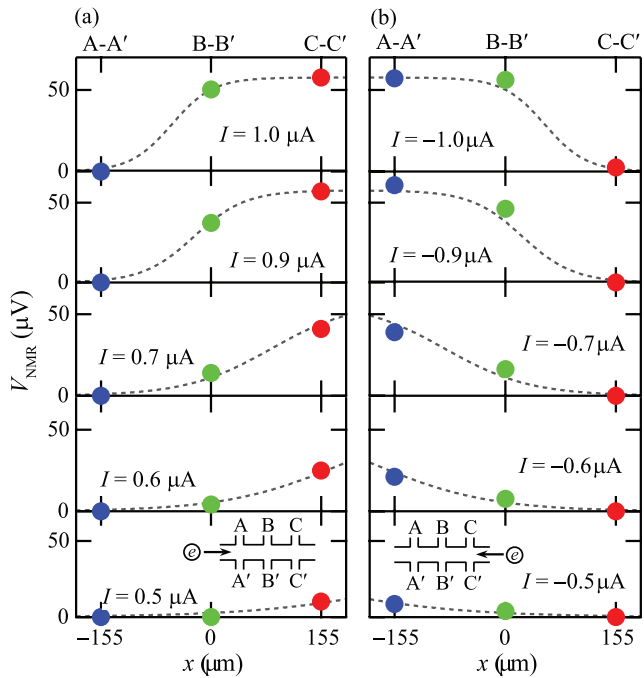}
	\caption{
		\label{probeposition}
		Amplitudes of the NMR signals obtained at various values of $I$ plotted
		against the distance $x$ measured from the center of the Hall bar. 
		The insets in the bottom panels of (a) and (b) indicate the electron-drift direction
		for $I > 0$ and $I < 0$, respectively.
		The dashed curves are visual guides.
		}
	\end{center}
\end{figure}

Figure~\ref{probeposition} shows the amplitudes of NMR signals $V_{\rm NMR}$ plotted
against the position $x$ of the Hall voltage probes measured from the center of the Hall bar.
Data obtained at various positive and negative values of $I$ are plotted
in Figs.~\ref{probeposition}(a) and (b), respectively.
At $I$ = 0.5 $\mu$A [bottom panel in Fig.~\ref{probeposition}(a)], 
a slight NMR signal is visible  only at probe C-C$^\prime$.
With increasing $I$, 
the amplitude of $V_{\rm NMR}$  increases at  probes B-B$^\prime$ and C-C$^\prime$.
The region where the NMR signal can be observed spreads toward the small $x$ region
with increasing $I$, as shown in Fig.~\ref{probeposition}(a).
Above $I$ = 0.9 $\mu$A, the $V_{\rm NMR}$ at probe B-B$^\prime$ reaches
a value close to that at probe C-C$^\prime$, 
and the $V_{\rm NMR}$-$x$ curve  saturates with increasing $x$.
When the polarity of $I$ is reversed ($I < 0$),
the amplitude of $V_{\rm NMR}$ decreases with increasing $x$ [Fig.~\ref{probeposition}(b)].
The $V_{\rm NMR}$-$x$ characteristics for each value of $|I|$ with opposite current polarities
are almost symmetric to each of their positive-polarity counterparts with respect to
the center of the Hall bar.
When the polarity of the magnetic field was reversed,
this pattern of $V_{\rm NMR}$-$x$ characteristics 
did not change significantly  (not shown).

Since the spatial pattern was changed by reversing the current polarity,
inhomogeneity in the 2DES is excluded
as the origin of the spatial dependence of $V_{\rm NMR}$.
The edge channel transport also does not play a significant role 
in the $V_{\rm NMR}$-$x$ characteristics because the characteristics
was not changed by reversing the magnetic field polarity.
Therefore, the observed spatial gradient of $V_{\rm NMR}$
is attributed to the spatially distributed nature of the DNP induced by QHE breakdown.
The observed dependence on current polarity dependence indicates that 
the amplitude of $V_{\rm NMR}$ increases as the distance 
from the electron-injecting electrode increases
and that the saturation of the $V_{\rm NMR}$-$x$ curve occurs in the region distant
from the electron-injecting electrode.
These features are similar to the spatial distribution of electron temperature $T_{e}$ discussed 
in  earlier studies of QHE breakdown at even-integer $\nu$\cite{Komiyama1996, Kaya1999}:
a gradual increase in $T_{e}$ over a macroscopic distance
along the electron-drift direction
and saturation of $T_{e}$ in a region sufficiently
distant  from the electron-injecting electrode [Fig.~\ref{IVcurve}(b)].
This similarity suggests the relevance
of the avalanche multiplication of excited electrons to the spatial distribution of DNP,
supporting our proposed scenario\cite{Kawamura2007}.
We think that  avalanche multiplication 
produces the spatial distribution not only of $T_{e}$
but also of electron spin flip events,
which in turn leads to the spatial distribution of  DNP through hyperfine interaction.

\begin{figure}[tbp]
	\begin{center}
	\includegraphics[width=7.5cm]{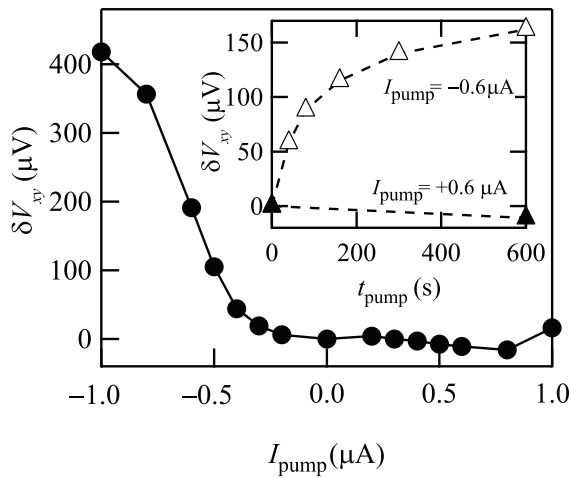}
	\caption{
		\label{pnp}
		Dependence of $\delta V_{xy}$ on $I_{\rm pump}$,
		obtained by the pump-and-probe measurement
			described in the text.
		The pump duration $t_{\rm pump}$ is fixed at 600 s.
		$\delta V_{xy}$ is defined 
		as $\delta V_{xy}$ = $V_{xy}(I_{\rm pump}, t_{\rm pump}) - V_{xy}(I_{\rm pump}, 0)$.
		The inset shows the dependence of $\delta V_{xy}$ on $t_{\rm pump}$ 
		for the cases of $I_{\rm pump}$ = $\pm$0.6 $\mu$A.
				}
	\end{center}
\end{figure}

Finally, in order to unambiguously confirm that
the absence of NMR signals at the voltage probe
near the electron-injecting electrode is not due to poor sensitivity of
$V_{xy}$ to the DNP,
we performed  pump-and-probe-type measurements similar
to those in  earlier studies\cite{Kawamura2008}.
DNP was induced by applying pumping current $I_{\rm pump}$
for a pumping time of $t_{\rm pump}$.
The value of $V_{xy}$ was measured at a probe current of
$I_{\rm probe}$ = 1.0 $\mu$A
 using Hall voltage probes A-A$^\prime$ before and after the pumping procedure. 
The change in $V_{xy}$ at $I_{\rm probe}$ = 1.0 $\mu$A,
$\delta V_{xy}(I_{\rm pump}) = V_{xy}(I_{\rm pump}, t_{\rm pump}) - V_{xy}(I_{\rm pump}, 0)$,
is induced by the DNP  pumping. 
As shown in the inset of Fig.~\ref{pnp},  $\delta V_{xy}$ increases
gradually with increasing $t_{\rm pump}$ when $I_{\rm pump}$ = $-$0.6 $\mu$A,
whereas the change in $\delta V_{xy}$ is negligibly small when $I_{\rm pump}$ = $+$0.6 $\mu$A.
This result suggests that nuclear spins are not polarized in the region
near the electron-injecting electrode
even for a pumping current $I_{\rm pump}$ larger than the $I_{\rm c}$ of QHE breakdown. 
Figure~\ref{pnp} shows the dependence of $\delta V_{xy}$ on  $I_{\rm pump}$
for $t_{\rm pump}$ = 600 s.
The value of $\delta V_{xy}$ increases only for 
$I_{\rm pump}  < $ $-$0.5 $\mu$A where electrons are 
injected from the electrode farther from the voltage probe. 
These results unambiguously show that the absence of $V_{\rm NMR}$
at probe A-A$^\prime$
 under the condition of $I > 0$ is attributed 
to the absence of DNP pumping, even for $I > I_{\rm c}$.

To summarize, 
we observe NMR signals in the breakdown regime of $\nu$ =1 QHE
by measuring Hall voltages using three different pairs 
of Hall voltage probes attached to the conducting channel of a Hall bar.
We find that the amplitude of the NMR signal becomes large as the distance
from the electron-injecting electrode increases.
Combined with  results on pump-and-probe measurements, 
we conclude that DNP induced by QHE breakdown has a spatial gradient
along the electron-drift direction in the Hall-bar device.
The similarity in the  spatial distribution between NMR signals 
and $T_{e}$ supports the scenario that the electron excitation between the 
spin-resolved Landau levels is relevant to the DNP.
In the case of the breakdown of odd-integer QHE,
both hyperfine and spin-orbit interactions cause the flips of electron spins
to excite them into the upper Landau subband with opposite spin polarity.
Given that only the contribution of the hyperfine interaction
is probed by the NMR measurements,
models of the QHE breakdown taking both spin-orbit and hyperfine interactions
into account are required for a quantitative understanding 
of the spatial profile of $V_{\rm NMR}$.
We think that studies on the spatial distribution of DNP will give further insight
into the mechanisms of DNPs reported under  various conditions of integer
fractional QH systems\cite{Kronmuller1999,Kawamura2009,Kou2010}.


\begin{thebibliography}{99}
\bibitem{fillingfactor}
The filling factor was intentionally detuned from $\nu$ = 1 because, at $\nu$ = 1,
the classical Hall coefficient $R_{\rm H}$ =1/$ne$ = $h/e^2$ and
$\Delta V_{xy}$ is expected to be zero even when $I > I_{\rm c}$.
\end{thebibliography}
\end{document}